\documentclass[aps,prl,twocolumn,showpacs]{revtex4}
\usepackage{graphicx}

\begin{document}

\title{Kondo temperature of magnetic impurities at surfaces}

\author{P. Wahl, L. Diekh\"oner, M.A. Schneider, L. Vitali, G. Wittich, and K. Kern}
\affiliation{Max-Planck-Institut f\"ur Festk\"orperforschung,
Heisenbergstr. 1, D-70569 Stuttgart, Germany}

\date{\today}

\begin{abstract}
Based on the experimental observation, that only the close vicinity of a magnetic impurity at metal surfaces determines its Kondo behaviour, we introduce a simple model which explains the Kondo temperatures observed for cobalt adatoms at the (111) and (100) surfaces of Cu, Ag, and Au. Excellent agreement between the model and scanning tunneling spectroscopy (STS) experiments is demonstrated. The Kondo temperature is shown to depend on the occupation of the $d$-level determined by the hybridization between adatom and substrate with a minimum around single occupancy.
\end{abstract}


\maketitle

Understanding the physics of a single spin supported on a metal host is at the basis of a bottom up approach to the design of high density magnetic recording \cite{Gambardella03}. The effects occuring in the limit of very small magnetic structures can be studied macroscopically on dilute magnetic alloys or microscopically for single magnetic impurities \cite{Madhavan98,Li98} and spins in quantum dots \cite{Goldhaber98,Cronenwett98}. These systems develop a rich phenomenology - which is commonly referred to as the Kondo problem \cite{Hewson93}. It deals with the interaction of a magnetic impurity with the conduction electrons of a surrounding non-magnetic metal host. This interaction leads to the screening of the spin of the impurity and in consequence to anomalies in the macroscopic properties. A many body ground state is formed at temperatures well below the Kondo temperature $T_\mathrm K$. Once the Kondo temperature of a given system is known, its behaviour at low temperatures is completely determined. The first experimental evidence became available 70 years ago by measurements of the resistivity of nonmagnetic metals with minute amounts of magnetic impurities \cite{Haas34} which showed an anomalous behaviour below $T_\mathrm K$. It is only recently, that interest has revived through the investigation of Kondo phenomena in quantum dots on one hand - providing model systems, which allow tuning of the relevant Kondo parameters easily \cite{Goldhaber98,Cronenwett98}, and of single magnetic impurities using low temperature scanning tunneling microscopy (STM) and spectroscopy (STS) on the other hand. The spectroscopic signature of the Kondo effect, the Kondo resonance, of single adatoms has been first observed by STS for cobalt adatoms on Au(111) \cite{Madhavan98} and cerium on Ag(111) \cite{Li98}.\\
The Kondo resonance shows up as a sharp peak in the local density of states (LDOS) which is pinned to the Fermi level and has a width proportional to $T_\mathrm K$. Since the first measurements on single magnetic impurities, several other Kondo systems at noble metal surfaces have been reported \cite{Jamneala00,Schneider02,Knorr02}. Although previously systematic studies of the Kondo effect of impurities on surfaces have been performed \cite{Jamneala00}, it is still unclear, what determines the Kondo behaviour of a specific adatom/substrate system.\\
In this letter, we present a comprehensive study of the Kondo resonance of single cobalt adatoms on different noble metal surfaces as well as on a heteroepitaxial system, namely Ag on Cu(111). Based on the observation, that the Kondo behaviour is governed by the immediate vicinity of the impurity, we introduce a simple model that describes the coupling between adatom and substrate and allows to predict the properties of the Kondo systems. The derived scaling behaviour is in perfect agreement with available experimental data for cobalt adatoms on noble metal surfaces.\\
The measurements have been performed in a home-built low temperature UHV-STM operating at $6\mathrm K$. The single crystal Cu(111) and Ag(100) surfaces have been carefully prepared by argon ion sputtering and annealing cycles in UHV (base pressure $1\cdot 10^{-10}\mathrm{mbar}$). Adatoms have been evaporated from a cobalt wire wound around a tungsten filament onto the surface at $\approx 20\mathrm K$. Typical coverages were around $0.001\mathrm{ML}$. The Ag monolayer has been deposited from an e-beam evaporator. Spectroscopic measurements were performed with open feedback loop using a lock-in technique with a modulation of the sample voltage of $1-1.5\mathrm{mV_{RMS}}$ at a frequency of $4.5 \mathrm{kHz}$. All bias voltages are sample potentials measured with respect to the tip and energies are given with respect to $E_\mathrm F$. Spectra shown in this letter are background subtracted \cite{Lauhon01}, which means that a spectrum taken on a clean spot of the surface is subtracted from the spectrum taken on the atom with the same tip. This removes artifacts due to slowly varying features in the tip spectrum. For the measurements presented here, an iridium tip, cut from a wire and mechanically ground, has been used. \\
The tunneling conductance with the tip placed on top of a magnetic adatom can be described by the well known Fano formula \cite{Fano61, Ujsaghy00}
\begin{equation}
\frac{\mathrm d I}{\mathrm d V}\sim \frac{(q+\tilde\epsilon)^2}{1+\tilde\epsilon^2},
\label{fano-function}
\end{equation}
where $\tilde\epsilon=\frac{\omega-\epsilon_\mathrm K}{\Gamma}$, $\omega=e\cdot V$ ($e$ is the elementary charge). It describes the lineshape due to the hybridization of the Kondo state with the conduction band electrons. The shape of the Fano function is determined by $q$ and can vary between a dip for $q=0$, an asymmetric lineshape for $q\approx 1$ and a peak for $q\rightarrow\infty$. $\epsilon_\mathrm K$ defines the position of the resonance relative to $E_\mathrm F$ and $\Gamma$ its half width. The latter is proportional to the Kondo temperature $T_\mathrm K$ \cite{ambig}. Although there has been considerable effort in describing the adatom/substrate system theoretically it is still unclear whether the lineshape as observed in STS experiments and hence $q$ is determined rather by interfering tunneling channels \cite{Plihal01} or by the electronic structure of the substrate \cite{Merino03,Lin03}.\\
\begin{figure}
 \includegraphics[width=8cm]{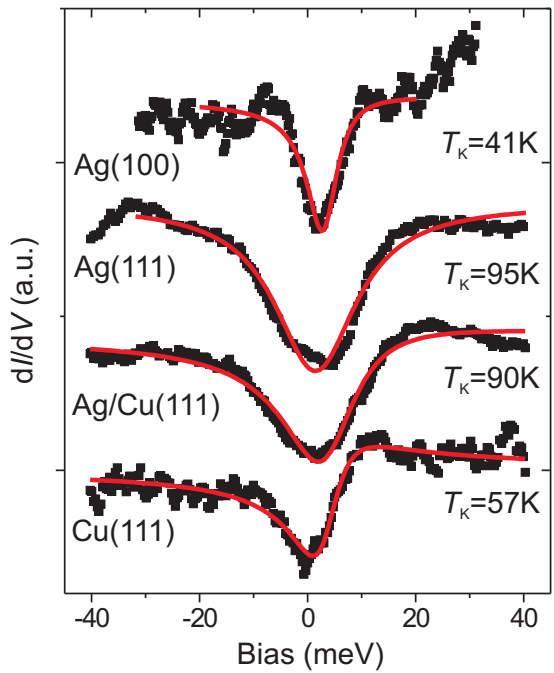}
 \caption{Spectra taken on a cobalt adatom on Ag(100) as well as on cobalt adatoms on Ag(111), one monolayer of Ag on Cu(111) and on Cu(111). The solid lines depict the fit of a Fano line shape (Eq.~\ref{fano-function}) to the data.}
 \label{fig1}
\end{figure}
\begin{table*}
\begin{tabular}{lllllllll}
\hline\hline
Substrate & $T_\mathrm K$ $[\mathrm K]$ & $q$ & $\epsilon_\mathrm K$ $[\mathrm{meV}]$ & $a$ $[\mathrm \AA]$ & $n_\mathrm{NN}$ & $\beta$ & $n_d$ & Ref. \\ \hline
Cu(111) & $54\pm2$ & $0.18\pm0.03$ & $1.8\pm0.6$ & $2.53$ & $3$ & $0.239$ & $0.94$ & \onlinecite{Knorr02} \\
& $53\pm5$ & $\cdots$ & $\cdots$ & $2.53$ & $3$ & $0.239$ & $0.94$ & \onlinecite{Manoharan00} \\
Cu(100) & $88\pm4$ & $1.13\pm0.06$ & $-1.3\pm0.4$ & $2.53$ & $4$ & $0.319$ & $1.16$ & \onlinecite{Knorr02} \\
Ag/Cu(111) & $92\pm10$ & $0.15\pm0.10$ & $2.3\pm1.0$ & $2.7$ & $3$ & $0.202$ & $0.84$ & this work \\
Ag(111) & $92\pm6$ & $0.0\pm0.1$ & $3.1\pm0.5$ & $2.7$ & $3$ & $0.202$ & $0.84$ & \onlinecite{Schneider02} \\
Ag(100) & $41\pm5$ & $0.0\pm0.2$ & $2.0\pm1.1$ & $2.7$ & $4$ & $0.269$ & $1.02$ & this work \\
Au(111) & $76\pm8$ & $\cdots$ & $\cdots$ & $2.695$ & $3$ & $0.203$ & $0.84$ & \onlinecite{Nick02} \\
& $75\pm6$ & $0.60\pm0.05$ & $6.5\pm0.5$ & $2.695$ & $3$ & $0.203$ & $0.84$ & \onlinecite{Madhavan01} \\
\hline\hline
\end{tabular}
\caption{Summary of Kondo temperatures for cobalt adatoms on the noble metal surfaces together with the parameters $a$ and $n_\mathrm{NN}$ which enter our model. $\beta$ is calculated according to Eq.~\ref{hop} and $n_d$ is obtained from the fit of Eqs.~\ref{scaling-n}-\ref{linnd} as described in the text.}
\label{table1}
\end{table*}
Typical tunneling spectra, which we measure with the tip placed above a single cobalt adatom are shown in Fig.~\ref{fig1}. For Ag(100), the spectrum shows a distinct dip slightly above the Fermi energy. The line shape is the same as for Co/Ag(111) \cite{Schneider02} while the Kondo temperature is $41\pm5\mathrm K$. The Kondo temperature has been averaged for several different atoms and tips. The value of the Kondo temperature for Co/Ag(100) is lower than that for Co/Ag(111). This indicates that the scaling of the Kondo temperature solely based on the number of nearest neighbours of the cobalt impurity as proposed for cobalt on copper surfaces \cite{Knorr02} has to be amended.\\
Experiments on heteroepitaxial monolayers do prove, however, that the Kondo behavior of surface impurities is essentially determined by the interaction with its nearest neighbors. To demonstrate this, we have deposited one monolayer of Ag on the Cu(111) substrate at room temperature and subsequently deposited single cobalt adatoms on top as described above. The spectra recorded are shown in Fig.~\ref{fig1}. The Kondo temperature of the adatom on one monolayer of silver on Cu(111) is already the same as on a bulk silver sample, although the Kondo temperatures of Ag and Cu differ by almost a factor of two. It is evident that the Kondo behaviour is governed by the interaction with the nearest neighbours of the impurity.  The Kondo temperatures as well as the lineshapes for Cobalt adatoms on noble metal surfaces are listed in table~\ref{table1}.\\
Based on the observation that the Kondo resonance is governed by local interactions we propose a model which explains the observed trend in the Kondo temperatures and the position of the resonance $\epsilon_\mathrm K$. In the Anderson model which describes the behaviour of a paramagnetic impurity in a nonmagnetic host metal the Kondo temperature $T_\mathrm K$ can be calculated from \cite{Hewson93}
\begin{equation}
k_\mathrm BT_\mathrm K\sim\sqrt{\frac{\Delta\cdot U}{2}}e^{-\frac{\pi}{2\Delta\cdot U}|\epsilon_d+U|\cdot |\epsilon_d|}.
\label{scaling}
\end{equation}
This equation relates the Kondo temperature $T_\mathrm K$ to the on-site Coulomb repulsion $U$, the half width of the hybridized $d$-level $\Delta$ and its position $\epsilon_d$. As can be easily seen, it is a monotonous function in $U$ and $\Delta$, while it has a minimum for $\epsilon_d=-\frac{U}{2}$ rising with either increasing or decreasing $\epsilon_d$. Related to the position $\epsilon_d$ is the shift of the Kondo resonance $\epsilon_\mathrm K$ with respect to the Fermi level. Due to level repulsion, the Kondo resonance is shifted to energies above $\epsilon_\mathrm F$ for $\epsilon_d>-\frac{U}{2}$ and to lower energies for $\epsilon_d<-\frac{U}{2}$. Table~\ref{table1} shows together with the Kondo temperatures the experimentally determined shift $\epsilon_\mathrm K$ of the Kondo resonance with respect to the Fermi level. The values of $\epsilon_\mathrm K$ indicate, that the position of the $d$-level varies considerably for the different adsorbate-substrate systems. It is positive for the (111)-surfaces, rising with the Kondo temperature when changing the substrate from Cu to Ag to Au. For the (100)-surfaces, the shift is increasing as well when going from Cu to Ag, but the Kondo temperature is decreasing - indicating, that the Kondo temperature has a minimum as a function of the position of the resonance. Closely related to the position $\epsilon_d$ and the on-site Coulomb repulsion $U$ is the occupation of the $d$-level $n_d$, which is larger than one for $\epsilon_d<-\frac{U}{2}$. Based on the experimental values of $\epsilon_\mathrm K$, one can readily establish a trend in  $\epsilon_d$ and $n_d$.\\
Following \'Ujs\'aghy {\it et al.} \cite{Ujsaghy00}, Eq.~\ref{scaling} can be expressed in terms of the occupation $n_d=-\frac{\epsilon_d}{U}+\frac{1}{2}$ of the $d$-level. Thus the properties of the adatom as derived from a single particle model can be mapped on an effective spin-$\frac{1}{2}$ Anderson model. This results in
\begin{equation}
k_\mathrm BT_\mathrm K\sim\sqrt{\frac{\Delta\cdot U}{2}}e^{-\frac{\pi U}{2\Delta}|-n_d+\frac{3}{2}|\cdot |-n_d+\frac{1}{2}|}.
\label{scaling-n}
\end{equation}
As discussed above the behavior of $T_\mathrm K$ can be understood in terms of the position of the $d$-level. Therefore we keep $\Delta$ and $U$ constant and use $\Delta=0.2\mathrm{eV}$ and $U=2.84\mathrm{eV}$ as calculated for Co/Au(111) by \'Ujs\'aghy {\it et al.} \cite{Ujsaghy00}. $n_d$ can vary between 0 (empty orbital) and 2 (double occupancy), where the Kondo regime is roughly $0.8<n_d<1.2$ \cite{Zlatic85} with approximately one unpaired electron in the $d$-level. \\
We assume, that the change in occupation of the $d$-level for cobalt adatoms and therefore $n_\mathrm d$ on the various noble metal surfaces discussed here can be estimated by a simple model which considers the hybridization between the $d$-orbitals of the impurity with the neighbouring substrate atoms. The occupation is assumed to increase with increasing overlap due to $sp-d$ hybridization \cite{Rodriguez94,Pacchioni99}. The hybridization of the $d$-orbital with the states of the substrate is described by a tight-binding-like hopping term \cite{Spanjaard82}
\begin{equation}
\beta=n_\mathrm{NN}e^{-\frac{a}{\lambda_d}}
\label{hop}
\end{equation} 
with a prefactor $n_\mathrm{NN}$ which takes the number of nearest neighbour substrate atoms into account. $a$ denotes the distance between the center of the adatom and of the next substrate atom which we calculate within a hard sphere model. The values used in the calculation are tabulated in table~\ref{table1}. The spatial extent of the $d$-orbital is assumed to be $\lambda_d\approx1\mathrm\AA$. The adatom resides in the hollow site on both types of surfaces discussed here, yielding $n_\mathrm{NN}=4$ for the (100)-surfaces and $n_\mathrm{NN}=3$ for the (111)-surfaces. To first order, $n_d$ is described in terms of the hybridization by
\begin{equation}
n_d=n_{d0}+c\cdot \beta.
\label{linnd}
\end{equation}
We use three fitting parameters in our model to explain the experimentally determined values of $T_\mathrm K$ and $\epsilon_\mathrm K$ as summarized in table~\ref{table1}: $n_{d0}$, $c$ and the proportionality constant in Eq.~\ref{scaling-n}. The resulting curve $T_\mathrm K(n_d)$ is shown in fig.~\ref{kondo-n}(a), where the Kondo temperature is plotted as a function of $n_\mathrm d$ as determined from eq. \ref{hop} and \ref{linnd}. It shows an excellent agreement with the experiments. The parameters extracted from the fit are $n_{d0}=0.28$, $c=2.78$ and for the proportionality constant $2.08$.
\begin{figure}
\includegraphics[width=9 cm]{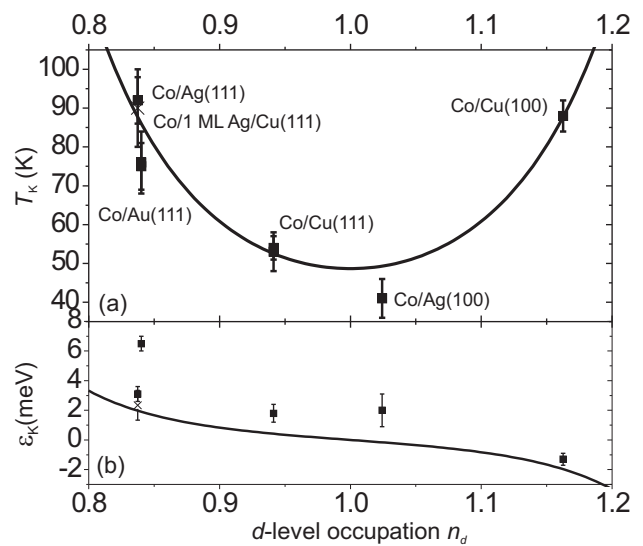}
\caption{(a) Kondo temperature $T_\mathrm K$ as function of the occupation $n_d$ of the $d$-level calculated with Eqs.~\ref{scaling-n}-\ref{linnd}. (b) Shift of the Kondo resonance $\epsilon_\mathrm K$ with respect to the Fermi energy, the solid line shows the prediction by our model according to Eq.~\ref{epskshift}.}
\label{kondo-n}
\end{figure}
As expected, the hybridization between the impurity and the conduction electrons and thus the occupation of the $d$-level varies in a narrow range around $n_d\approx1$. The trend in the occupation of the $d$-level is confirmed by the position $\epsilon_\mathrm K$ of the Kondo resonance. The relation between $n_d$ and $\epsilon_\mathrm K$ is plotted in Fig.~\ref{kondo-n}(b) together with the theoretically expected behaviour \cite{Hewson93}
\begin{equation}
\epsilon_\mathrm K=\Gamma\tan\left(\frac{\pi}{2}(1-n_d)\right).
\label{epskshift}
\end{equation}
The systems cobalt on Cu(111) and Ag(100) are almost perfectly described by the symmetric Anderson model, where $n_d=1$ and the Kondo temperature as a function of the occupation has a minimum. Either by enhancing the hybridization, i.e. by changing the substrate to Cu(100), or by reducing it by going to Ag(111) or Au(111) the Kondo temperature is rising. The value, which we find for the occupation of the $d$-level for Co/Au(111), 0.84 is close to the calculated value 0.8 \cite{Ujsaghy00}. The physical picture that emerges is that for non-integer occupation the spin flip probability is higher and thus the Kondo temperature increases.\\
The key result is that based on simple arguments one finds an increasing occupation of the impurity level expressed in the value of $n_\mathrm d$ when going from a Ag(111) substrate to Cu(100) which explains the observed values of $T_\mathrm K$ and $\epsilon_\mathrm K$ using the mapping onto the Anderson model proposed in Ref.~\cite{Ujsaghy00}. The limitations of the simple model are obvious. It neglects relaxation, which can lead to an enhanced hybridization between the $d$-level and the substrate. Furthermore, details of the electronic structure of the substrates are ignored. For Co/Au(111) the Kondo temperature is slightly lower than predicted by our model. This fact reflects that the model does not take into account hybridization with specific substrate orbitals. While the $4d$-orbitals of silver are very localized and well below the fermi energy, the $5d$-orbitals of Au are close to the fermi level and will contribute to the hybridization with the impurity. Thus the overlap is most likely underestimated for gold and overestimated for silver. Finally, the model neglects that an enhanced hybridization will lead to a reduction of the Coulomb repulsion $U$ and an increase in the width of the $d$-level $\Delta$. $T_\mathrm K$ increases monotonically as a function of both smaller $U$ and larger $\Delta$. Taking the variation of both into account will render the curve shown in fig.~\ref{kondo-n}(a) asymmetric and its minimum will be shifted to lower occupations but qualitatively show the same trend. This explains the systematic deviation of the shift $\epsilon_\mathrm K$ in fig.~\ref{kondo-n}(b) from the calculated curve.\\
We note that the surface state does not play a major role. Although its onset on Ag(111) is $165\mathrm{meV}$ closer to $E_\mathrm F$ than on Ag/Cu(111) \cite{Wessendorf04}, the Kondo temperature of a cobalt adatom is the same. This is consistent with our model, where the properties of the surface state do not enter and with earlier experimental \cite{Knorr02, Schneider02} and theoretical results \cite{Barral04} which both show that the surface state has only a minor influence on the properties of Kondo systems. \\
The behaviour of the Kondo temperature as shown in fig.~\ref{kondo-n}(a) is similar to what has been measured on quantum dots. There, the properties of the Kondo system are tuned by varying a gate voltage, which shifts the electronic states up or down and thus decreases or increases the fractional occupation of the quantum dot \cite{Cronenwett98}. In the case of adatom/surface systems, this tuning can be achieved by modifying the substrate.\\
In conclusion, we have presented a simple model in order to understand the large range of Kondo temperatures of cobalt adatoms on noble metals surfaces which have been reported previously and in this work. The model estimates the hybridization between the adatom's $d$-level and the substrate. Despite the simplicity of this model, which is in contrast to the complexity of the many body physics necessary to describe the Kondo effect, it captures the essence of the impurity/substrate Kondo systems presented here. Motivated by experiments on Ag/Cu(111), only the local environment of the impurity enters our model. The Kondo temperatures for adatoms on multilayer systems is shown to depend only on the topmost layer. This conclusion is not trivial, since the extent of the Kondo cloud - the size of the many body state which leads to the formation of the Kondo resonance - is believed to be on the order of several ten to hundred nanometers \cite{Affleck96}. Due to the local character of the magnetic interaction, it will be interesting to investigate the behaviour of a Kondo impurity on a surface alloy as substrate, which should allow for a fine tuning of the Kondo temperature as a function of the local stoichiometry in the vicinity of the adatom. \\
We gratefully acknowledge fruitful discussion with O. Gunnarsson, J. Merino and H. Kroha.\\


\begin{references}
\bibitem{Gambardella03} P. Gambardella {\it et al.}, Science {\bf 300}, 1130 (2003).
\bibitem{Madhavan98} V. Madhavan, W. Chen, T. Jamneala, M.F. Crommie, and N.S. Wingreen, Science {\bf 280}, 567 (1998).
\bibitem{Li98} J. Li, W.-D. Schneider, R. Berndt and B. Delley, Phys. Rev. Lett. {\bf 80}, 2893 (1998).
\bibitem{Goldhaber98} D. Goldhaber-Gordon {\it et al.}, Nature {\bf 391}, 156 (1998).
\bibitem{Cronenwett98} S.M. Cronenwett, T.H. Oosterkamp and L.P. Kouwenhoven, Science {\bf 281}, 540 (1998).
\bibitem{Hewson93} A.C. Hewson, {\it The Kondo Problem to Heavy Fermions} Cambrigde (1993).
\bibitem{Haas34} W.J. de Haas, J. de Boer and G.J. van den Berg, Physica {\bf 1}, 1115 (1934).
\bibitem{Knorr02} N.Knorr, M.A. Schneider, L. Diekh\"oner, P. Wahl, and K. Kern, Phys. Rev. Lett. {\bf 88}, 096804 (2002).
\bibitem{Schneider02} M.A. Schneider, L. Vitali, N. Knorr, and K. Kern, Phys. Rev. B {\bf 65}, 121406 (2002).
\bibitem{Jamneala00} T. Jamneala, V. Madhavan, W. Chen, and M.F. Crommie, Phys. Rev. B {\bf 61}, 9990 (2000).
\bibitem{Lauhon01} L.J. Lauhon and W. Ho, Rev. Sci. Instr. {\bf 72}, 216 (2001).
\bibitem{Fano61} U. Fano, Phys. Rev. {\bf 124}, 1866 (1961).
\bibitem{Ujsaghy00} O. \'Ujs\'aghy, J. Kroha, L. Szunyogh, and A. Zawadowski, Phys. Rev. Lett. {\bf 85}, 2557 (2000).
\bibitem{Manoharan00} H.C. Manoharan, C.P. Lutz, and D.M. Eigler, Nature {\bf 403}, 512 (2000).
\bibitem{Nick02} N. Knorr, PhD Thesis, Lausanne (2002).
\bibitem{Madhavan01} V. Madhavan, W. Chen, T. Jamneala, M.F. Crommie, and N.S. Wingreen, Phys. Rev. B {\bf 64}, 165412 (2001).
\bibitem{ambig} There is some ambiguity in the definition of the Kondo temperature \cite{Hewson93}.
\bibitem{Plihal01} M. Plihal and J.W. Gadzuk, Phys. Rev. B {\bf 63}, 085404 (2001).
\bibitem{Merino03} J. Merino and O. Gunnarsson, Phys. Rev. B {\bf 69}, 115404 (2004).
\bibitem{Lin03} Ch.-Y. Lin, A.H. Castro Neto, B.A. Jones, cond-mat/0307185 (unpublished).
\bibitem{Zlatic85} V. Zlati\'c, B. Horvati\'c and D. \v Sok\v cevi\'c, Z. Phys. B {\bf 59}, 151 (1985).
\bibitem{Pacchioni99} G. Pacchioni, M. Mayer, S. Kr\"uger and N. R\"osch, Chem. Phys. Lett. {\bf 299}, 137 (1999).
\bibitem{Rodriguez94} J.A. Rodriguez, Surf. Sci. {\bf 303}, 366 (1994).
\bibitem{Spanjaard82} M.C. Desjonqu\`eres and D. Spanjaard, J. Phys. C, {\bf 15}, 4007 (1982).
\bibitem{Wessendorf04} M. Wessendorf {\it et al.}, Appl. Phys. A {\bf 78}, 183 (2004).
\bibitem{Barral04} M. A. Barral, A.M. Llois and A.A. Aligia, Phys. Rev B {\bf 70}, 035416 (2004).
\bibitem{Affleck96} E.S. S\o rensen and I. Affleck, Phys. Rev. B {\bf 53}, 9153 (1996).
\end{references}
\end{document}